\documentstyle[aps,prl,epsf,floats]{revtex} 
\draft

\begin{document}
\twocolumn[\hsize\textwidth\columnwidth\hsize\csname
@twocolumnfalse\endcsname 


\preprint{UCLA/00/TEP/28 
} 

\title{Possible galactic sources of ultrahigh-energy cosmic rays and a
strategy for their detection via gravitational lensing }

\author{Alexander Kusenko$^{1,2}$ and Vadim A. Kuzmin$^3$ }
\address{$^1$Department of Physics and Astronomy, UCLA, Los Angeles, CA
90095-1547, USA \\ $^2$RIKEN BNL Research Center, Brookhaven National
Laboratory, Upton, NY 11973, USA \\ $^3$Institute for Nuclear Research,
Russian Academy of Sciences, 60th October Anniversary Prosp. 7a, Moscow
117312, Russia }

\date{December 1, 2000}

\maketitle
             
\begin{abstract}
If decays of superheavy relic particles in the galactic halo are
responsible for ultrahigh-energy cosmic rays, these particles must be
clustered to account for small-scale anisotropy in the AGASA data.  We show
that the masses of such clusters are large enough for them to gravitationally
lens stars and galaxies in the background.  We propose a general
strategy that can be used to detect such clusters via gravitational
lensing, or to rule out the hypothesis of decaying relic particles as the
origin of highest-energy cosmic rays.
 
\end{abstract}

\pacs{PACS numbers: 98.70.S, 95.75.D  \hspace{1.0cm}
BNL-HET-00/43; UCLA/00/TEP/28} 

\vskip2.0pc]

\renewcommand{\thefootnote}{\arabic{footnote}}
\setcounter{footnote}{0}

The origin of cosmic rays~\cite{data,data1,clustering} with energies beyond
the Greisen-Zatsepin-Kuzmin (GZK) cutoff~\cite{gzk} is unknown.  One of the
possible explanations invokes decays of metastable superheavy relic
particles $X$ with masses $10^{13}$GeV or higher and cosmologically long
lifetimes~\cite{particles,kr,y}.  Such superheavy particles could be produced
non-thermally at the end of inflation~\cite{ckr,kt,kt1}.  Their extremely
small decay width may be due to a conservation of some topological
charge~\cite{kr}. Particles with the requisite properties may also arise
from string theory~\cite{cryptons}.

If these particles decay into hadrons and photons, the flux of
ultrahigh-energy cosmic rays (UHECR) is dominated by those particles in the
halo of our galaxy~\cite{particles}.  This can explain the absence of the
GZK cutoff.  Even if the superheavy particles decay predominantly into
neutrinos~\cite{gk2}, cosmic rays with energies beyond the GZK cutoff may
originate through Z-bursts~\cite{Zburst}.  In this {\em letter}
we concentrate on the former possibility and assume that observed
ultrahigh-energy events~\cite{data,data1,clustering} come mainly from the
decays of relic particles in the Milky Way halo.

The new data provide an opportunity to test this hypothesis through
gravitational lensing.  There is a strengthening evidence for directional
clustering of events in the AGASA data~\cite{data1,clustering}.  The latest
analyses~\cite{data1} show one triplet and six doublets of events, each
originating from the same point in the sky, to $\pm 1.3^\circ$ accuracy.
The probability of this clustering to occur by accident is less than
$0.07\%$~\cite{data1}.  The only way to reconcile these data with the
hypothesis of relic particle decays is to assume a non-uniform distribution
of particles in the halo.  If the relic particles form regions of increased
density, such lumps may be responsible for the doublets in the UHECR data.
To produce a doublet, a lump of particles must be of a certain size
determined by the decay probability. Since the mass of the hypothetical
particle is fixed by the energy of UHECR, there is a prediction for the
mass of each lump that can give rise to a doublet.  In addition, the
celestial coordinates of the particle cluster are known to one degree
accuracy.  In this {\em letter} we propose a novel gravitational lensing
technique that can be used to discover a cluster of relic particles, or to
rule out such particles as the origin of ultrahigh-energy cosmic rays.

Under the assumption that UHECR are caused by the relic particle decays,
the data suggest that (in addition to a possible uniform distribution), of
the order of ten clusters of X-particles exist in our galactic halo.
N-body simulations of dark matter halos predict some inhomogeneities that
can be related to small-scale anisotropy of UHECR~\cite{blasi} but probably
are not sufficient to produce larger clumps.  However, additional
interactions of the hypothetical particles can alter this picture
dramatically. We assume that each doublet comes from a separate cluster of
particles.  Here we do not discuss the dynamics of clustering of the heavy
particles. This issue will be addressed in an upcoming publication.  We
note in passing that clumps of dark matter with masses $10^8M_\odot$ may
resolve~\cite{clumps_cusps} the widely debated issue of cusps in the halo
density profiles~\cite{steinhardt}.

If the X-particle lifetime is $\tau_{_X}\sim 10^{10}-10^{22} 
{\rm yr}$~\cite{kr}, a cluster of $N$ particles produces decays at a 
rate 
\begin{equation}
P= \frac{N}{\tau_{_X}} \sim (10^{-10}-10^{-22}) N \ {\rm yr}^{-1} .
\end{equation}

The probability for the decay products to produce an air shower in a
detector is $P\times (d/L)^2 $, where $d\sim 10^6$~cm is the size of the
detector and $L\sim 10^{23}$~cm is the distance to the cluster of relic
particles.  In order to have a doublet in a one-year data set, each
cluster must have
\begin{equation}
N\sim \frac{\tau_{_X}}{1 \ {\rm yr}} \left ( \frac{L}{d} \right )^2  
\sim  10^{50}  \left ( \frac{\tau_{_X}}{10^{16}\ {\rm yr}} \right ) 
\end{equation}
particles.  If X-particle has mass $m_{_X}\sim
10^{13}$GeV~\cite{particles,kr}, the mass of the cluster is 
\begin{equation}
M \sim 10^{63} \left ( \frac{\tau_{_X}}{10^{16}\ {\rm yr}} \right ) \ {\rm
GeV}= 10^{6} \left ( \frac{\tau_{_X}}{10^{16}\ {\rm yr}} \right ) \ M_\odot .
\end{equation}
The lifetime $\tau_{_X}$ can be in the range from $10^{10}$~yr (for the
relic particles to survive until present) to $10^{21}$~yr (for the total
mass of the clusters not to exceed the mass of the galaxy).
Correspondingly, the masses of clusters can range from one solar mass to
$10^{10} M_\odot $. 
 
There is a remarkable possibility to discover such invisible massive
objects by what we will call a ``lens-chasing'' technique. 
Although the cluster can be entirely dark, it can be detected
through gravitational lensing of stars and galaxies behind it.  AGASA
data~\cite{data,data1} provide the celestial coordinates of the clusters
with a precision of a few degrees.

\begin{figure}[t]
\centering
\hspace*{-5.5mm}
\leavevmode\epsfysize=8.5cm \epsfbox{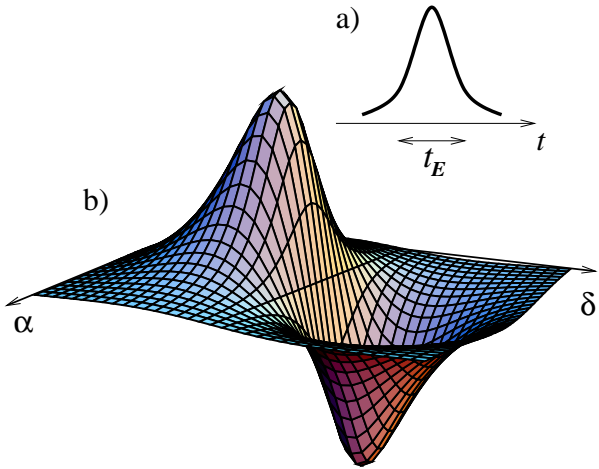}\\[3mm]
\caption[fig1]{\label{fig1} A typical light curve for gravitational lensing
of a single source as a function of time (a), and the corresponding plot of
brightness variations for a sample of stars (b) as a function of celestial
coordinates, in arbitrary units.  Photometric changes indicate the
location of the lens. }
\end{figure}

The Einstein radius of a cluster with mass $M$ is 

\begin{equation}
R_{_E} = 0.14 \; {\rm pc} \left [ \frac{M}{10^{7} M_\odot} \frac{L}{10 \,
{\rm kpc}} \right ]^{1/2}.
\end{equation}
A lens of this kind has angular size 

\begin{equation}
\theta_{_E} = \frac{R_{_E}}{L} = 0.0008^\circ \left [ \frac{M}{10^{7}
M_\odot} \frac{10 \, {\rm  kpc}}{L} \right ]^{1/2} 
\end{equation}
and passes the line of sight in time 

\begin{equation}
t_{_E} = \frac{R_{_E}}{10^{-3}c}= 4.8 \times 10^2 \ {\rm yr} \ \left [
\frac{M}{10^{7} M_\odot} \frac{L}{10 \, {\rm kpc}} \right ]^{1/2} .
\label{dt}
\end{equation}

We propose to use the small changes in brightness of stars and galaxies
behind the lens on time scales of the order of one year.  The AGASA data
specify the location of the cluster to $\pm 1.3^\circ$.  It is possible to
scan over a large sample of remote sources in the patch specified by the
UHECR data, recording the brightness of the background stars and galaxies.
The scan must be repeated after a period of several months.  Next, one
should extract the changes in the absolute brightness of the background
stars.  A slowly moving lens with a single-star light curve shown in
Fig.~1(a) will produce a map of brightness differentials shown in Fig.1(b).
By using temporal changes in the brightness of stars from a large sample,
one can locate a small lens within a large, $2.5^\circ$, patch of the sky.

The number of background sources chosen for scanning and photometry 
determines the sensitivity of the proposed lens-chasing experiment.  Let us
consider a sample of $n^2$ stars with an average angular separation of
$\theta_b \approx 2.5^\circ/n$.  A lens that passes near one of these stars
at an angular distance $\delta \theta = \theta_b/2 <\theta_{_E} $ from the
line of sight will magnify the source by a factor $(2 \theta_E/\theta_b)^2
$ as compared to its brightness in the absence of the lens.  The change in
the star's brightness over a one year period is
\begin{eqnarray}
\frac{\Delta A}{A} & \approx & \frac{1 \; {\rm yr}}{t_{_E}} \left ( \frac{
\theta_E}{\theta_b/2} \right )^2 =  
\frac{1 \; {\rm yr}}{t_{_E}} \left ( \frac{2 n 
\theta_E}{2.5^\circ } \right )^2 \nonumber \\ 
& = & 0.8 \%  \; \left ( \frac{n^2}{10^7} \right )  
\left [ \frac{M}{10^{7} M_\odot} \right ]^{1/2} \left [
\frac{10 \, {\rm  kpc}}{L} \right ]^{3/2}  .
\end{eqnarray}

Assuming a better than $1\%$ precision photometry, a sample of $10^7$
background sources allows detection of clusters with mass $10^7 M_\odot$
and higher.  Smaller masses require a higher number of background
sources. For comparison, MACHO project has monitored 11.9 million stars
during 5.7 years of operation~\cite{macho}.  Of course, lens chasing
presents a very different challenge from that faced by MACHO.  Unlike
MACHO, which monitors bright nearby stars in Large Magellanic Cloud (LMC)
on a continuous basis, we want a relatively infrequent (once a year)
accurate photometry of stars in the directions of UHECR doublets.

Some of the clusters in the AGASA data lie in the supergalactic
plane~\cite{data1,clustering}.  The presence of many relatively close (and,
hence, bright) background stars in these directions makes the corresponding
clusters particularly appealing for lens chasing.

One can, of course, refine this technique.  If the lensing
cluster is discovered after several initial crude scans, one can narrow
down its coordinates and perform a more detailed monitoring of closely
spaced sources around the location of the lens.

We note in passing that future detectors can observe yet another signature
of the same kind of sources.  Decays of superheavy particles and subsequent
fragmentation can produce excited hadrons.  Their decays, in turn, can
produce simultaneous air showers separated by thousands of kilometers.  The
time delay is $\delta t = t \gamma^{-2} \Delta E/E $, where $t$ is the
time of flight, $\gamma \sim 10^{11}$ is the Lorentz factor, and $\Delta
E/E \sim 1$.  The difference in the arrival time $\delta t \sim 10^{-11}$s,
and the distance between air showers is of the order of $10^3$~km.
Future space-based detectors, such as EUSO and OWL, can observe such
spatially separated events in coincidence. 

To summarize, a small-scale anisotropy in the AGASA data demands that, if
the UHECR are due to decaying relic particles in the halo, these particles
form clusters with coordinates specified by the cosmic ray events.  The
masses of such clusters can range from $M_\odot$ to $10^{10}M_\odot$.  We
have proposed a general strategy for detecting such clusters by their
gravitational lensing of the background stars and galaxies.

This work was supported by the NATO Collaborative Linkage Grant
PST.CLG.976397.  In addition, A.K. was supported by the US Department
of Energy, grant DE-FG03-91ER40662, Task C.  The work of V.A.K. was
supported in part by the grant 98-02-1744a from Russian Foundation for
Basic Research.  V.A.K. thanks UCLA for hospitality during his visit, when
this work was performed.


\end{document}